\documentclass[prb,twocolumn,showpacs,preprintnumbers,amsmath,amssymb]{revtex4}

\usepackage{graphicx}
\graphicspath{{fig/}}
\usepackage{dcolumn}
\usepackage{bm}
\usepackage{amsmath,amssymb}
\usepackage[T1]{fontenc}
\usepackage{textcomp}
\usepackage{color}

\begin{document}

\title{Josephson and Persistent Spin Currents in Bose-Einstein Condensates of Magnons
}

\author{Kouki Nakata,$^{1,2}$ 
Kevin A. van Hoogdalem,$^1$  Pascal Simon$^3$    and Daniel Loss$^{1}$}

\affiliation{$^1$Department of Physics, University of Basel,   Klingelbergstrasse 82, CH-4056 Basel, Switzerland   \\
$^2$Yukawa Institute for Theoretical Physics, Kyoto University, Kitashirakawa, Kyoto 606-8502, Japan  \\
$^3$Laboratoire de Physique des Solides, CNRS UMR-8502, Univ. Paris Sud, 91405 Orsay Cedex, France}

\date{\today}

\begin{abstract}
Using the Aharonov-Casher (A-C) phase,
we present a microscopic theory of the Josephson and persistent spin currents in quasi-equilibrium Bose-Einstein condensates (BECs) of magnons in ferromagnetic insulators.
Starting from a microscopic spin model that we  map onto a Gross-Pitaevskii  Hamiltonian, we derive a two-state model for the Josephson junction between the weakly coupled magnon-BECs.  We then show how to obtain the alternating-current (ac) Josephson effect with magnons as well as macroscopic quantum self-trapping in a magnon-BEC. 
We next propose how to control the direct-current (dc) Josephson effect electrically using the A-C phase, which is the geometric phase acquired by magnons moving in an electric field. 
Finally,  we introduce a magnon-BEC ring and show that persistent magnon-BEC currents flow due to the A-C phase. 
Focusing on the feature that the persistent magnon-BEC current is a steady flow of magnetic dipoles that produces an electric field, we propose a method to directly measure it experimentally.
\end{abstract}


\maketitle

\section{Introduction}
\label{sec:intro}

One of the urgent issues in modern electronics is the removal of waste energy due to Joule heating. Spintronics \cite{awschalom,mod} offers a possible solution to this problem by replacing charge currents  with (pure) spin currents. \cite{magnon2}
 Therefore, establishing methods to generate spin currents, and understanding the transport of magnetic excitations are significant both from the viewpoint of fundamental science as well as applications to information- and communication technologies.

Traditionally, spintronics mainly concerned itself with the transport of spin-polarized conduction electrons in metallic magnets. A standard way to generate a spin current in such spin-polarized metals is by means of spin pumping.\cite{silsbee,ISHE1,battery,mod2,takeuchi,nakatatatara,QSP,QSPtemp} The resulting spin current can be indirectly measured using the inverse spin-Hall effect.\cite{ISHE1} Although this is an experimentally established method, the crucial issue is that the pumped spin current in metals disappears within a very short distance (typically a few micrometers). This has been an obstacle to the practical use of spin currents in metallic systems so far.

Recently, Kajiwara $\textit{et al}$.\cite{spinwave} have experimentally addressed this issue by employing spin currents in ferromagnetic insulators. Although such insulators\cite{TserkovnyakNatNano,bender,Takei,wei} contain no conduction electrons, they do host a qualitatively different type of magnetic carrier, namely  magnons (or spin wave excitations).
These excitations describe the collective motion of the exchange-coupled spins of the localized electrons in ferromagnetic insulators.
It is also worth pointing out that in such  non-itinerant systems, the dissipation problem is reduced since
true magnetization transport generates typically much
less power than charge currents.\cite{Trauzettel}
In their experiment, Kajiwara $\textit{et al}$.\cite{spinwave} showed that it is possible to electrically create and read-out a spin-wave spin current in the magnetic insulator Y$_3$Fe$_5$O$_{12}$ (YIG) using spin-Hall effects. Furthermore, they determined that the resulting spin-wave spin current can be transmitted over distances of several millimeters, much further than what is typically possible when using spin-polarized conduction electrons in magnetic metals.

\begin{figure}[h]
\begin{center}
\includegraphics[width=8cm,clip]{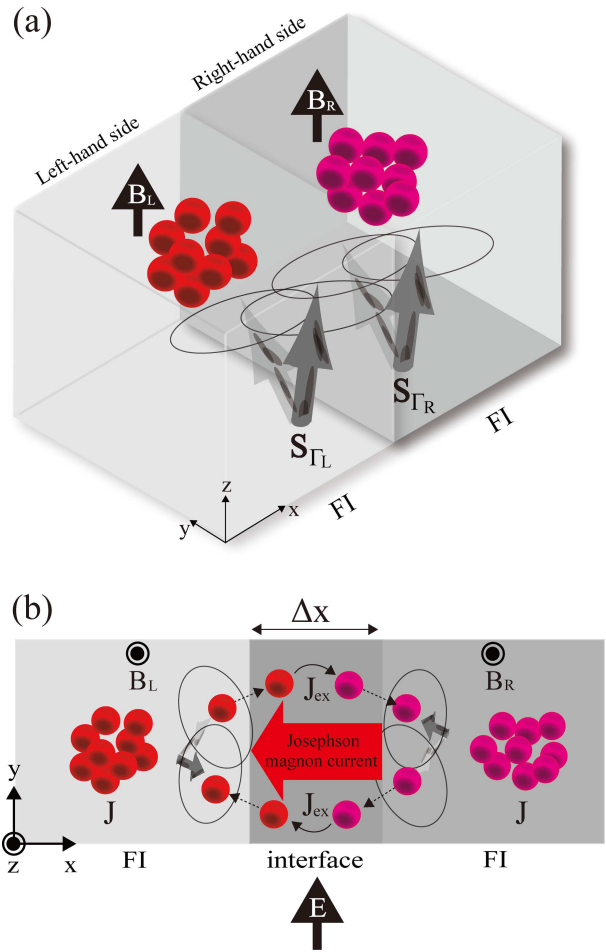}
\caption{(Color online)
(a) Schematic representation of the quasi-equilibrium magnon-BEC Josephson junction (MJJ) consisting of two coupled ferromagnetic insulators (FI)  in the presence of magnetic fields ${\bf B}_{\rm L}$ and ${\bf B}_{\rm R}$.
Each cloud of circles represents a single magnon-BEC. A quasi-equilibrium magnon-BEC corresponds to a macroscopic coherent precession of the spins in each FI. 
The boundary spins ${\bf S}_{\Gamma_{\rm L}}$ in the left FI and ${\bf S}_{\Gamma_{\rm R}}$ in the right FI are relevant to the Josephson magnon current.
(b) Close-up of the MJJ. The two FIs are separated by an interface of width $\Delta x$ and thereby weakly exchange-coupled with strength $J_\textrm{ex}$. 
The variable $J $ represents the exchange interaction between neighboring spins in each FI.
The applied electric field, ${\mathbf{E}} = E{\bf e}_y $, couples to the magnons through the A-C phase.
 \label{fig:Josephson} }
\end{center}
\end{figure}

An additional advantageous property of magnons lies in their bosonic nature, which qualitatively distinguishes them from fermionic conduction electrons. One of the consequences of this bosonic nature is that magnons can form a Bose-Einstein condensate (BEC).\cite{BatistaBEC} Indeed, Demokritov $\textit{et al}$.\cite{demokritov} have experimentally shown that the quasi-equilibrium magnon-BEC\footnote{Regarding the semantic issue about the terminology BEC,  see  Refs. \onlinecite{BatistaBEC,kloss,snoke,bunkov,sonin}.} can be produced even at room temperature in YIG by using microwave pumping.

Stimulated by the experimental progress achieved by Kajiwara $\textit{et al}$.\cite{spinwave} and Demokritov $\textit{et al}$.,\cite{demokritov} we theoretically propose an alternative method to generate and control spin-wave spin currents  (referred to as magnon currents) in ferromagnetic insulators using quasi-equilibrium magnon-BECs.\cite{demokritov}  
To this end we study both the alternating current (ac) as well as the direct current (dc) Josephson effects in junctions consisting of weakly coupled magnon-BECs (see Fig. \ref{fig:Josephson}),
in analogy to the original Josephson effects in superconductors. \cite{Josephson} For earlier work on Josephson effects in magnetic systems see Refs. \onlinecite{ytDL,TataraJosephson,Schilling,troncoso2}.
 We determine the effect of the Aharonov-Casher (A-C) phase\cite{casher} on the ac and dc Josephson effects, and find that the A-C phase gives us a handle to control the different Josephson effects using electric fields.  
Finally, we introduce a magnon-BEC ring.
We then show that persistent\cite{LossPersistent,LossPersistent2} magnon-BEC currents flow in the ring due to the A-C phase.
Focusing on the feature that the persistent magnon-BEC current is a steady flow of magnetic dipoles (\textit{i.e.} magnons), we illustrate how to directly measure it \cite{dipole,magnon2} experimentally.

This paper is organized as follows: in Sec. \ref{sec:system}, we introduce the model we will use to describe a quasi-equilibrium magnon-BEC Josephson junction (MJJ). In Sec. \ref{sec:ACJosephson} we discuss the ac Josephson effect as well as macroscopic quantum self-trapping (MQST) in a quasi-equilibrium magnon-BEC. Then, in Sec. \ref{sec:weak} we present our results on the dc Josephson effect and determine the effect of an electric field applied to the MJJ. 
Finally, we introduce a magnon-BEC ring in Sec. \ref{sec:ring}. We show that persistent magnon-BEC currents flow due to the A-C phase. A method to directly measure the steady flow of magnetic dipoles (\textit{i.e.} magnons) is also proposed at the end of this section.

\section{Magnon-BEC Josephson junction}
\label{sec:system}

\subsection{Gross-Pitaevskii Hamiltonian}
\label{subsec:pascal}

We consider the setup depicted in Fig. \ref{fig:Josephson}. It consists of two weakly exchange-coupled ferromagnetic insulators (FIs). We will assume that the FIs are identical, but different magnetic fields ${\bf B}_{\rm{L(R)}}(t) = B_{\rm{L(R)}} (t) {\bf e}_z$ are applied to respectively the left- and right FI [$ {\mathbf{e}}_{\nu } $ denotes the unit vector along the $\nu $-axis ($\nu = x, y, z$)]. Furthermore, we apply an electric field ${\bf E} = E {\bf e}_y$ to the interface. We are interested in the transport of the Bose-Einstein condensed magnons between the two FIs as a function of these different applied fields.

 A microscopic spin Hamiltonian  $  {\cal{H}}_{\rm{H}}  $  that describes a single FI is given by
\begin{eqnarray}
    {\cal{H}}_{\rm{H}}  =  \sum_{\langle i j\rangle} {\bf S}_i \cdot {\bf J} \cdot {\bf S}_j -  g\mu_{\rm{B}}  {\bf B} \cdot \sum_i {\bf S}_i,
 \label{eqn:Heisenberg}
\end{eqnarray}
where ${\bf J}$ denotes a diagonal $3\times 3$-matrix with $\textrm{diag}({\bf J}) = J\{ 1, 1,\eta\}$. The variable $J <0$ represents the exchange interaction between neighboring spins in the ferromagnetic insulator and  $  \eta >0 $ denotes the anisotropy of the spin Hamiltonian. We assume a cubic lattice and a magnetic field ${\bf B}(t) = B(t) {\bf e}_z$ is applied. By using the Holstein-Primakoff\cite{HP} transformation, $ S_i^+ = \sqrt{2S}[1-a_i^\dagger a_i / (2S)]^{1/2} a_i$, $S_i^z = S - a_i^\dagger a_i$, we can map Eq. (\ref{eqn:Heisenberg}) onto a system of magnons: chargeless bosonic quasi-particles with magnetic moment $g\mu_{\rm{B}}  {\bf e}_z$. Magnons are described by creation/annihilation operators $a_{i}^{(\dagger)}$ that satisfy the commutation relation $[a_{i}, a_{j}^{\dagger }] = \delta_{i,j}$.

We will assume that the magnons in both FIs have undergone quasi-equilibrium Bose-Einstein condensation, through a procedure such as realized in Ref. \onlinecite{demokritov}. This implies that the magnon state in either FI is given by a coherent macroscopic quantum state characterized by a finite expectation value $\langle a_i \rangle$. Since we are interested in the semiclassical dynamics of the quasi-equilibrium magnon-BEC, we take the continuum limit in Eq. (\ref{eqn:Heisenberg}) and replace the operator $ a_{i}$ with its expectation value in the coherent state, given by  $\langle a({\bf r}, t) \rangle = \sqrt{n ({\bf r}, t)} {\rm{exp}}[i \vartheta  ({\bf r}, t)]  $. The variable $n  ({\bf r}, t) \equiv  \langle a({\bf r}, t)\rangle^{\ast } \langle a ({\bf r}, t) \rangle$ represents the number density of magnons and $\vartheta   ({\bf r}, t)$ denotes the phase.

To make the connection to the standard theory of Josephson effects in BECs, we rewrite Eq. (\ref{eqn:Heisenberg}) in terms of the parameters $n({\bf r}, t), \vartheta ({\bf r}, t)$. After some straightforward manipulations, we obtain the well-known\cite{Zapata} Gross-Pitaevskii (GP) Hamiltonian $  {\cal{H}}_{\rm{GP}} $ given by
\begin{eqnarray}
{\cal{H}}_{\rm{GP}} &=& \int   \textrm{d} {\bf r}
                                        \Big{\{}  g_1 \left[ \left|\nabla  \sqrt{n } \right|^2 + n  \left|{\nabla {\vartheta  }}\right|^2   \right]   \nonumber    \\
                                     &+& g_2  {n}^{2}    + B_{\rm{eff}}  n  \Big{\}}.
 \label{eqn:GP}
\end{eqnarray}
The parameters of the GP Hamiltonian are given in terms of the original parameters of the microscopic model as
\begin{eqnarray}
 g_1  = - JS (1  + \eta  ) \alpha  ^2,        \     \    
 g_2  =  -J  (1  - \eta  )    \alpha ^3,   
 \label{eqn:gp2}
\end{eqnarray}
and
\begin{eqnarray}
      B_{\rm{eff}} = 4 JS  (1   -   \eta  )  +g\mu_{\rm{B}}  B.
       \label{eqn:gp3}
\end{eqnarray}
The parameter $\alpha $ describes the lattice constant. From Eqs. (\ref{eqn:GP}) and (\ref{eqn:gp2}) we recover then the expected result that the system is noninteracting  in the isotropic case (i.e. $ \eta  =1$), whereas $\eta \lessgtr 1$ corresponds to attractive (repulsive) magnon-magnon interactions.

\subsection{Aharonov-Casher Effect }
\label{subsec:A-C}
Next, we focus on the interface region connecting the two FIs. We will denote by ${\bf S}_{\rm{\Gamma_{L(R)}}}$ the localized spins that reside on that relevant two-dimensional boundary surface of the left (right) insulator. Due to a finite overlap of the wave functions,  there exists in general a finite exchange interaction between the spins located on the boundaries of the different FIs. Let us denote by $J_{\rm{ex}}$ ($ > 0 $) such energy scale. In the tunnel limit, one has therefore $J_{\rm{ex}}\ll  \mid J \mid $.
The  exchange interaction between the two FIs may be described by the Hamiltonian
\begin{equation}
 {\cal{H}}_{\rm{ex}} = -J_{\rm{ex}} \sum_{\langle \Gamma_{\rm{L}} \Gamma_{\rm{R}} \rangle} {\bf S}_{\Gamma_{\rm{L}}} \cdot {\bf S}_{\Gamma_{\rm{R}}}.
\end{equation}
In terms of magnon operators, this interaction can be rewritten as $  {\cal{H}}_{\rm{ex}} = -J_{\rm{ex}} S  \sum_{\langle \Gamma_{\rm{L}} \Gamma_{\rm{R}} \rangle}  (a_{\Gamma_{\rm{L}}} a_{\Gamma_{\rm{R}}}^{\dagger } + a_{\Gamma_{\rm{L}}}^{\dagger } a_{\Gamma_{\rm{R}}})  $.
We have ignored terms arising from the $z$-component of the spin variables in the Hamiltonian ${\cal{H}}_{\rm{ex}}$, since these do not influence the dynamics of the junction in any significant manner.

Finally, we focus on the effect of an applied electric field on the transport of the magnons. A magnetic dipole $g\mu_{\rm{B}}  {\bf e}_z$ moving along a path $\gamma$ in an electric field ${\bf E}({\bf r})$ acquires a geometric phase
\begin{equation}
 \theta _{\rm{A{\mathchar`-}C}} = \frac{g \mu_{\rm{B}}}{\hbar c^2} \int_{\gamma} \textrm{d} {\bf l} \cdot \left[{\bf E}({\bf r}) \times  {\bf e}_z\right].
  \label{eqn:acac}
\end{equation}
This phase is called the Aharonov-Casher phase.\cite{casher} It is dual to the Aharonov-Bohm phase.\cite{bohm} Both geometric phases are a special case of a Berry phase.\cite{Mignani,Hea} The tunneling Hamiltonian  $ {\cal{H}}_{\rm{ex}} $ is changed when an electric field  $  {\mathbf{E}} = E{\bf e}_y $ is applied to the interface. It becomes\cite{magnon2}
\begin{eqnarray}
{\cal{H}}_{\rm{ex}}^{\rm{A{\mathchar`-}C}}  &=& -J_{\rm{ex}} S  \sum_{\langle \Gamma_{\rm{L}} \Gamma_{\rm{R}} \rangle}
                                                 (a_{\Gamma_{\rm{L}}} a_{\Gamma_{\rm{R}}}^{\dagger }  {\rm{e}}^{- i  \theta _{\rm{A{\mathchar`-}C}}}
                                              + {\rm{h.c.}}),
 \label{eqn:ac}
\end{eqnarray}
where $  \theta _{\rm{A{\mathchar`-}C}} = [g \mu_{\rm{B}}/(\hbar c^2)] E \Delta x$ for the geometry under consideration [see Fig. \ref{fig:Josephson} (b)].

\subsection{Two-State Model}
\label{subsec:twostate}
Next, we determine the time evolution of the MJJ under the Hamiltonian $ {\cal{H}}_{\rm{MJJ}} =  {\cal{H}}_{\rm{GP}}  + {\cal{H}}_{\rm{ex}}^{\rm{A{\mathchar`-}C}} $. 
To this end, we assume ${\cal{H}}_{\rm{ex}}^{\rm{A{\mathchar`-}C}}$ is a small perturbation and use the semi-classical approximation introduced in the previous section.
Using Hamilton's equation of motion we then find the effective two-state model\cite{smerzi} of the MJJ (Fig. \ref{fig:Josephson}):
\begin{subequations}
\begin{eqnarray}   
   i \hbar  \dot{\psi}_{\rm{L}} &=&           E_{\rm{L}}  \psi_{\rm{L}}   +  U_{\rm{L}}   n_{\rm{L}}  \psi_{\rm{L}}    - K_{\rm{L}} \psi_{\rm{R}},    \label{eqn:twostate}  \\
      i \hbar  \dot{\psi}_{\rm{R}} &=&           E_{\rm{R}}  \psi_{\rm{R}}   +  U_{\rm{R}}   n_{\rm{R}}  \psi_{\rm{R}}     - K_{\rm{R}} \psi_{\rm{L}},     
 \label{eqn:twostate2}
\end{eqnarray}
\end{subequations}
with 
\begin{subequations}
\begin{eqnarray} 
  E_{\rm{L(R)}} & = & 4 JS  (1   -   \eta  )  +g\mu_{\rm{B}}  B_{\rm{L(R)}},  \label{eqn:E} \\
 U_{\rm{L(R)}} &=& -2J(1-\eta) \alpha^3,    \label{eqn:U}    \\        
  K_{\rm{L}} &=& K_{\rm{R}}^{\ast } = K_{0}{\rm{e}}^{ i  \theta _{\rm{A{\mathchar`-}C}}}.
 \label{eqn:K}
\end{eqnarray}
\end{subequations}
Here we have replaced the operators $a_{\Gamma_{\rm L(R)}}$ in $ {\cal{H}}_{\rm{MJJ}}$ by their respective expectation values 
$\psi _{\rm L(R)}({\bf r}, t)=  \sqrt{ n_{\rm L(R)}} e^{i \vartheta_{\rm L(R)} } $ as before,
where  $ n_{\rm L(R)} = \psi_{\rm L(R)}^{\ast }\psi_{\rm L(R)}$ represents the number density of magnons in each ferromagnetic insulator and  $ \vartheta_{\rm L(R)} $ the phase. 
We have defined the tunneling amplitude $K_0  \equiv  J_{\rm ex} S$. Note that $  K_{\rm{L(R)}} $ becomes complex in the presence of an electric field. Here, the fact that the magnon eigenstates in a single FI are known (since the magnon-BEC state corresponds to the condensation of the magnons in the lowest momentum state) allows us to write the parameters $U_{\rm L(R)},E_{\rm L(R)},K_{\rm L(R)}$ in terms of the parameters of the original microscopic spin model.

We define the magnon population imbalance $z(t) \equiv  [n_{\rm{L}}(t) - n_{\rm{R}}(t)]/n_{\rm T}$ and the relative phase $\theta(t) \equiv \vartheta _{\rm R} (t)  -  \vartheta _{\rm L} (t)$. The constant  $n_{\rm T} \equiv  n_{\rm L}(t) + n_{\rm R}(t)$ denotes the total population in the MJJ. In terms of the canonically conjugate variables $z(t)$ and $ \theta (t)$, Eqs. (\ref{eqn:twostate})-(\ref{eqn:twostate2}) become
\begin{subequations}
\begin{eqnarray}
       \frac{ d z }{d \tau }   &=&  -   \sqrt{1- z^2} {\rm{sin}} (\theta + \theta _{\rm{A{\mathchar`-}C}}),     \label{eqn:twostate5}    \\ 
       \frac{ d \theta  }{d \tau }   &=&   \Delta  E + \Lambda  z  
                                             +    \frac{ z}{\sqrt{1-z^2}}  {\rm{cos}} (\theta + \theta _{\rm{A{\mathchar`-}C}}).
   \label{eqn:twostate6}
\end{eqnarray}
\end{subequations}
We have rescaled the time by  $ t \to \hbar  \tau/ 2 K_{0}  $, and have defined
\begin{subequations}
\begin{eqnarray}
       \Delta  E  &=& \frac{E_{\rm{L}}- E_{\rm{R}}}{2 K_{0}} + \frac{U_{\rm{L}}-U_{\rm{R}}}{4 K_0} n_{\rm{T}}, \label{eqn:parameter}     \\
       \Lambda  &=&  \frac{U_{\rm{L}} + U_{\rm{R}}}{4 K_0} n_{\rm{T}}.
   \label{eqn:parameter2}
\end{eqnarray}
\end{subequations}
Eqs. (\ref{eqn:twostate5}) and (\ref{eqn:twostate6}) are the fundamental equations of the Josephson effects in a MJJ in the presence of an electric field. From now on, on the basis of the Josephson equation shown by eqs. (\ref{eqn:twostate5}) and (\ref{eqn:twostate6}), we discuss the contribution of the A-C phase to the Josephson effects in quasi-equilibrium magnon-BEC and clarify the conditions for the ac and dc Josephson effects to occur. We note that $\Delta \mu\equiv  \dot{\theta}(\tau)$ is usually referred to as chemical potential difference.\cite{Levy}

Finally, let us remark that when $  \theta _{\rm{A{\mathchar`-}C}} =   0$, the description of the MJJ is mathematically identical to that of a Bose Josephson junction of atomic BECs.\cite{smerzi,smerzi2,smerzi3}
Similar equations for $  \theta _{\rm{A{\mathchar`-}C}} =   0$ have been proposed phenomenologically for antiferromagnets.\cite{Schilling}

\section{ac Josephson effect  and macroscopic quantum self-trapping}
\label{sec:ACJosephson}
\begin{figure}[h]
\begin{center}
\includegraphics[width=8.5cm,clip]{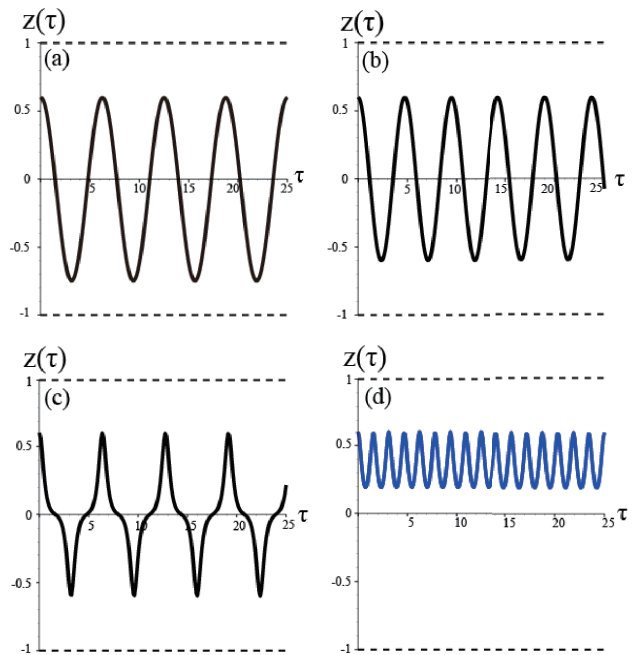}
\caption{(Color online)
The ac Josephson effect: plots of the population imbalance $z(\tau)$ as function of the rescaled time $\tau= (2  J_{\rm{ex}} S /\hbar) t $ obtained by numerically solving Eqs. (\ref{eqn:twostate5})-(\ref{eqn:twostate6})  for the values (a) 
$ \Delta E = 0.1 $,   $ \Lambda =0 $, $  \theta _{\rm{A{\mathchar`-}C}}   =0 $, $ z(0) =0.6 $, and $ \theta  (0) =0  $.
As an example, for $ J_{\rm{ex}} = 0.25$ $\mu$eV, $S=2$, and $ g=2 $, this corresponds to $  B_{\rm{L}} -  B_{\rm{R}} = 1 $ mT and the period of an oscillation is $ {\cal{T}} =  6$ ns ($\textit{i.e.}$ the rescaled time $ \tau =1 $ corresponds to $t = 1$ ns). 
(b) - (d) show  cases with vanishing 
 $ \Delta E =  \theta _{\rm{A{\mathchar`-}C}}  = 0 $, $ z(0) =0.6 $, and $ \theta  (0) =0  $,
which give $ \Lambda _{\rm{c}}^{\rm{MJJ}}   = 10$.
(b) $ \Lambda =1$,
(c) $ \Lambda =9.99$, and
(d) $ \Lambda =11 $.
When the value of $ \Lambda $ exceeds $  \Lambda_{\rm{c}}^{\rm{MJJ}} $, 
$  \Lambda  > \Lambda_{\rm{c}}^{\rm{MJJ}}$, the MQST occurs as shown in (d).
 \label{fig:MQST} }
\end{center}
\end{figure}

The ac Josephson effect is most easily understood by considering Eqs. (\ref{eqn:twostate5})-(\ref{eqn:twostate6}) with $\Lambda = 0$ (corresponding to isotropic ferromagnets) and $z\ll 1$. Applying a static magnetic field $B_{\rm L}$ to the left condensate leads to a uniformly increasing phase difference $\theta (t)$,
which in turn results in an ac magnon current through the junction.  The population imbalance for the ac Josephson effect is shown in Fig. \ref{fig:MQST} (a) for realistic experimental parameters. It is seen that an applied magnetic field of approximately 1 mT leads to oscillations with a frequency in the GHz-range.
By analogy with the original Josephson effect, we note that the magnon ac Josephson effect could be used in metrology to convert an applied magnetic field  
into a frequency.

Next, we focus on the phenomena of macroscopic quantum self-trapping (MQST)\cite{albiez,troncoso2} in our MJJ. MQST is defined by an oscillation of the population of magnons in each magnon-BEC around a non-zero time-averaged value, {\it i.e.} $ \overline{z (\tau)} \neq 0$: in the remainder of this section, we will put $ \Delta  E= 0$ and $\Lambda \neq 0$,  and focus on the effect of the A-C phase on the occurrence of MQST. We will consider the initial conditions $ z(0) \not=0  $ and $ \theta (0)=0 $. An initial population imbalance $ z(0) \not=0  $ could be realized by tuning the magnitude of the microwave\cite{demokritov} applied to each ferromagnetic insulator when generating the quasi-equilibrium magnon-BEC differently.

We find that MQST occurs\cite{smerzi2} when the value of the self-interaction $\Lambda$ satisfies $  \Lambda  > \Lambda_{\rm{c}}^{\rm{MJJ}}$, where 
\begin{eqnarray}
         \Lambda _{\rm{c}}^{\rm{MJJ}} 
= \frac{ 1+  \sqrt{1-z(0)^2}    {\rm{cos}} ( \theta(0) + \theta _{\rm{A{\mathchar`-}C}})   }{ z(0)^2/2   }.
   \label{eqn:critical}
\end{eqnarray}
Experimentally, controlling $\Lambda$ will probably most easily be achieved through the total number of magnons $n_T$ in the system. The onset of MQST as a function of $\Lambda$ is shown in Figs. \ref{fig:MQST} (b)-(d). It also follows from Eq. (\ref{eqn:critical}) that the application of an electric field, which leads to an increased  $\theta _{\rm{A{\mathchar`-}C}}$, reduces the value of $\Lambda$ required to observe the MQST effect. We note that when $ \theta _{\rm{A{\mathchar`-}C}} = 0$, the value of $\Lambda _{\rm{c}}^{\rm{MJJ}}  $ is reduced to the value found by Smerzi $\textit{et al.}$\cite{smerzi,smerzi2,smerzi3} in their study on Josephson effects in atomic BECs.

\section{dc Josephson effect in magnon-BECs}
\label{sec:weak}

Next, we focus on the dc Josephson effect\cite{Levy} in the presence of an A-C phase. We consider a junction consisting of two identical anisotropic ferromagnetic insulators with $\eta > 1$. We will show that it is possible to generate a dc Josephson magnon current by applying a time-dependent magnetic field to the (say) left FI such that the term $\Delta E(\tau)$ in the Josephson equation  (\ref{eqn:twostate6}) can be written as
\begin{equation}
     \Delta E(\tau) = \frac{g \mu_{\rm{B}} }{2 K_0} \left(B_{\rm{L}}  -B_{\rm{R}}\right) = \left\{ \begin{array}{cc} - b_0 \tau & \textrm{ for } \tau \in (0,\tau_0), \\ 
0 & \textrm{ otherwise }. \end{array} \right.
   \label{eqn:70}
\end{equation}
In other words, we apply a magnetic field whose magnitude increases over time with a rate proportional to $  b_0$, for a limited (renormalized) time $\tau_0$. 

To illustrate how this leads to the occurrence of the dc Josephson effect, we focus initially on the dynamics in the limit where $|z|\ll 1$ and $\Lambda \gg 1$. The latter corresponds to $J_{\rm ex} \ll  \mid  J \mid $, so we are in the weak-tunneling regime. Eqs. (\ref{eqn:twostate5}) and (\ref{eqn:twostate6}) can then be approximated as
\begin{subequations}
\begin{eqnarray}
       \frac{ d z }{d \tau }   &\approx &  - {\rm{sin}} (\theta  +\theta_{\rm{A{\mathchar`-}C}}),    \label{eqn:initial3}  \\
       \frac{ d \theta  }{d \tau }   &\approx &   - {b}_0 \tau + \Lambda z.
   \label{eqn:initial4}
\end{eqnarray}
\end{subequations}
It is immediately seen that the presence of a dc Josephson  magnon current is described by the steady-state solution
\begin{equation}
z (\tau) = {z} _0 \tau \textrm{ and } \theta(\tau) = -\arcsin({z}_0) - \theta_{\rm{A{\mathchar`-}C}},   \label{eqn:KoukiKevin}
\end{equation}
with ${z} _0 \equiv   {b}_0  / \Lambda$. We see from Eq. (\ref{eqn:KoukiKevin}) that the system can only sustain the (dissipationless) dc Josephson magnon current for values ${b}_0 \leq \Lambda$. Hence, the magnetic field must be increased adiabatically, \textit{i.e.} slowly enough such that enough magnons can tunnel through the junction to keep the difference in chemical potential $\Delta \mu = \dot{\theta}(\tau)$ equal to zero, in order for the dc Josephson effect to occur in this setup. If, on the other hand, the magnetic field is increased with a rate ${b}_0 > \Lambda$, a chemical potential difference across the junction builds up, and the system ends up in the ac regime.

\begin{figure}[h]
\begin{center}
\includegraphics[width=8.5cm,clip]{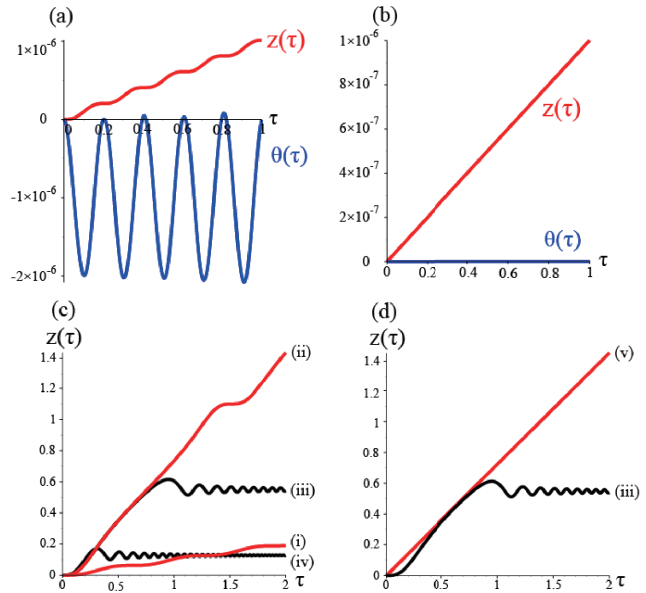}
\caption{(Color online)
The dc Josephson effect through the A-C phase: plots of the population imbalance $z(\tau)$ and the relative phase $\theta(\tau)$ as function of the rescaled time $\tau= (2  J_{\rm{ex}} S /\hbar) t $ obtained by numerically solving Eqs. (\ref{eqn:initial3})-(\ref{eqn:initial4})  for the values 
(a) (b)  $ {z} _0 =  10^{-6}  $, $ \Lambda =  10^{3} (\gg  1)$, and $z(0)=\theta (0) = 0$. 
(a) $\theta _{\rm{A{\mathchar`-}C}}  = 0$, in which the period of the small oscillation is estimated as $ {\cal{T}} \approx  0.2 $ ns.
(b) $ \theta _{\rm{A{\mathchar`-}C}}  =  -\arcsin({z}_0)=-10^{-6}$.
(c)
The breakdown of the dc Josephson effect ($\theta_{\rm{A{\mathchar`-}C}} =0$) due to increase of $b_0$. The transition between the dc region [(i), (ii)] and the ac one [(iii), (iv)] takes place for  $ {z} _0 \approx    0.725 $ due to the absence of the A-C phase. Under the conditions $(z(0), \theta (0)) =(0, 0)$ and $  \Lambda =100 $  ($ \gg  1 $), each $ {z} _0 = {b}_0  / \Lambda $  reads (i) 0.100, (ii) 0.724, (iii) 0.726, and (iv) 1.100. (d) Example of the recovery of the dc Josephson effect from the ac effect through the A-C phase in the region $ {z} _0 \leq 1 $   ({\textit{i.e.}} $ {b}_0  \leq  \Lambda $).
(iii) $\theta_{\rm{A{\mathchar`-}C}} =0$, (v) $\theta_{\rm{A{\mathchar`-}C}} = -\arcsin({z}_0)$.
 \label{fig:DCbreakdownACphase} }
\end{center}
\end{figure}

It should be stressed that the steady-state solution by itself does not necessarily completely determine the physics of the dc Josephson effect. Assuming the junction is in equilibrium at  $\tau = 0$, the initial phase difference is given by $\theta(0) = 0$. Unless $\theta_{\rm{A{\mathchar`-}C}}$ is tuned to the value
\begin{equation}
\theta_{\rm{A{\mathchar`-}C}} = -\arcsin({z}_0),  \label{eqn:A-CforDC}
\end{equation}
a mismatch in $\theta(\tau)$ with the steady-state solution arises. This mismatch leads to two related phenomena. Firstly, it leads to oscillations in both $z(\tau)$ and $\theta(\tau)$, as can be seen in Fig. \ref{fig:DCbreakdownACphase} (a) as opposed to the case  $\theta_{\rm{A{\mathchar`-}C}} = -\arcsin({z}_0)$ represented in  Fig. \ref{fig:DCbreakdownACphase} (b). However, even if $ {\rm d} \theta(\tau)/ {\rm d}\tau \neq 0$ for certain $\tau$, the time-average $ \overline{{\rm d} \theta /{\rm d} \tau} $ does still satisfy $  \overline{{\rm d}\theta /{\rm d} \tau} = 0$ for small enough $ {z}_0$. In that sense, the dc Josephson magnon current is robust against initial phase differences. On the other hand, we can view the fact that the oscillations can be tuned by $\theta_{\rm{A{\mathchar`-}C}}$ as one of the signatures of the A-C phase (albeit one that will be hard to demonstrate experimentally, given the prohibitively short period of the oscillations, see caption of Fig. \ref{fig:DCbreakdownACphase} (a)).

The second consequence of the mismatch due to the initial conditions is that the value of $ {z}_0$ where the transition between the dc- and the ac-regime occurs is reduced by a numerical factor $\approx 0.725$. This is illustrated in Fig. \ref{fig:DCbreakdownACphase} (c). Interestingly, this allows one in principle to cross the transition between the ac- and dc-regime for values of ${z}_0\gtrsim 0.725$ by tuning the A-C phase between $\theta_\textrm{A-C} = 0$ and $\theta_\textrm{A-C} \approx - \arcsin({z}_0)$, see Fig. \ref{fig:DCbreakdownACphase} (d). 

So far, we have neglected the influence of the cosine term in Eq. (\ref{eqn:twostate6}) on the occurrence of the dc Josephson effect. We have actually solved numerically the  system in Eqs. (\ref{eqn:twostate5})-(\ref{eqn:twostate6}) for different values of $\Lambda$. We have found that the effect of smaller values of $\Lambda$ (\textit{i.e.} an increased effect of the cosine term) is to increase the amplitude of the oscillations in $z(\tau), \theta(\tau)$. However, besides that, we check that the main mechanism behind the occurrence of the dc Josephson magnon current remains in place.

\begin{figure}[h]
\begin{center}
\includegraphics[width=7cm,clip]{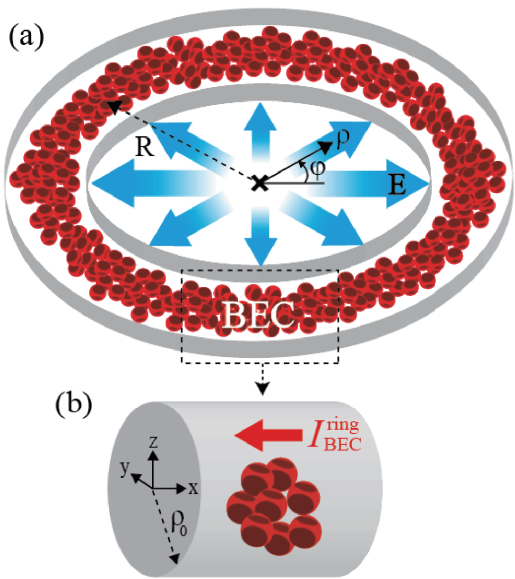}
\caption{(Color online)
(a) A schematic picture of the magnon-BEC ring. 
The radius of the ring reads $R$ and the applied electric field is ${\bf E}(\rho , \varphi ) =  {\cal{E}}  \rho  {\bf e}_{\rho }$.
(b) An enlarged view of the cylindrical wire that forms the ring.  The radius of the wire is $ \rho_0 $.
The persistent magnon-BEC current $   I_{\rm{BEC}}^{\rm{ring}}  $ flows in the ring due to the A-C phase.
\label{fig:SQUID} }
\end{center}
\end{figure}

\section{Persistent Magnon-BEC Current}
\label{sec:ring}

Lastly, based on the duality between the Aharonov-Bohm phase and the A-C phase, we introduce a magnon-BEC ring [Fig.  \ref{fig:SQUID} (a)], in analogy to a superconducting ring.\cite{SCring,ImryPersistent} 
Due to the A-C phase, persistent magnon-BEC currents flow in the ring. 
We remark that persistent spin currents in magnetic systems  have been considered before,\cite{Loss_Maslov,Kopietz,sigrist} but to the best
of our knowledge not yet for quasi-equilibrium magnon-BECs.\cite{demokritov,BatistaBEC}

\subsection{Magnon-BEC current}
\label{subsec:BECcurrent}

Let us now calculate the general expression for the magnon-BEC current.
For this we first consider a spin chain along the $x$-axis described by (see also Sec. \ref{subsec:A-C})
${\cal{H}}_{\rm{chain}}^{\rm{A{\mathchar`-}C}}  
=  J  S  \sum_{ j }   (a_{j} a_{j + 1}^{\dagger }  {\rm{e}}^{- i  \theta _{\rm{A{\mathchar`-}C}}^{\rm{chain}}} + \rm{h.c.}) $.
The A-C phase reads $  \theta _{\rm{A{\mathchar`-}C}}^{\rm{chain}} = [g \mu_{\rm{B}}/(\hbar c^2)] E \alpha  $,
where $E$ is the applied electric field  along the $y$-axis.
We assume that the magnons have undergone quasi-equilibrium Bose-Einstein condensation as before, through a procedure such as realized in Ref. \onlinecite{demokritov}. 
A uniform magnetic field is also applied along the  $z$-axis ({\textit{i.e.}} quantization axis), but we have ignored such terms arising from the $z$-component of the spin variables since these commute with the number operators of magnons on each site and they do not influence the dynamics of the magnon-BEC current in any significant manner.
The operator of the magnon-BEC current that flows from the $l$-site to ($l+ 1$)-site becomes 
$ I_{\rm{BEC}} = - g \mu_{\rm{B}} [N_{l},   J  S (a_{l} a_{l + 1}^{\dagger }  {\rm{e}}^{- i  \theta _{\rm{A{\mathchar`-}C}}^{\rm{chain}}} + {\rm{h.c.}})]/(i \hbar )$, where $N_{l} \equiv   a_{l}^{\dagger }a_{l}$ is the number operator of condensed magnons on the $l$-site.
A quasi-equilibrium magnon-BEC corresponds to the macroscopic coherent precession in terms of the original spins.
We then  treat it semiclassically and replace the operators $a_{l(l+ 1)}$ by their expectation values  as before (see Sec. \ref{subsec:pascal} and \ref{subsec:twostate}). 
Thus, the magnon-BEC current in the spin chain ${\cal{H}}_{\rm{chain}}^{\rm{A{\mathchar`-}C}}  $ becomes 
\begin{equation}
 I_{\rm{BEC}} = - 2 g \mu_{\rm{B}} (J/\hbar)  S N_{\rm{BEC}}  {\rm{sin}} \theta _{\rm{A{\mathchar`-}C}}^{\rm{chain}} \, ,
\label{eqn:BECcurrent}
\end{equation}
where $N_{\rm{BEC}}$  is the number of condensed magnons on each site.

\begin{figure}[h]
\begin{center}
\includegraphics[width=5cm,clip]{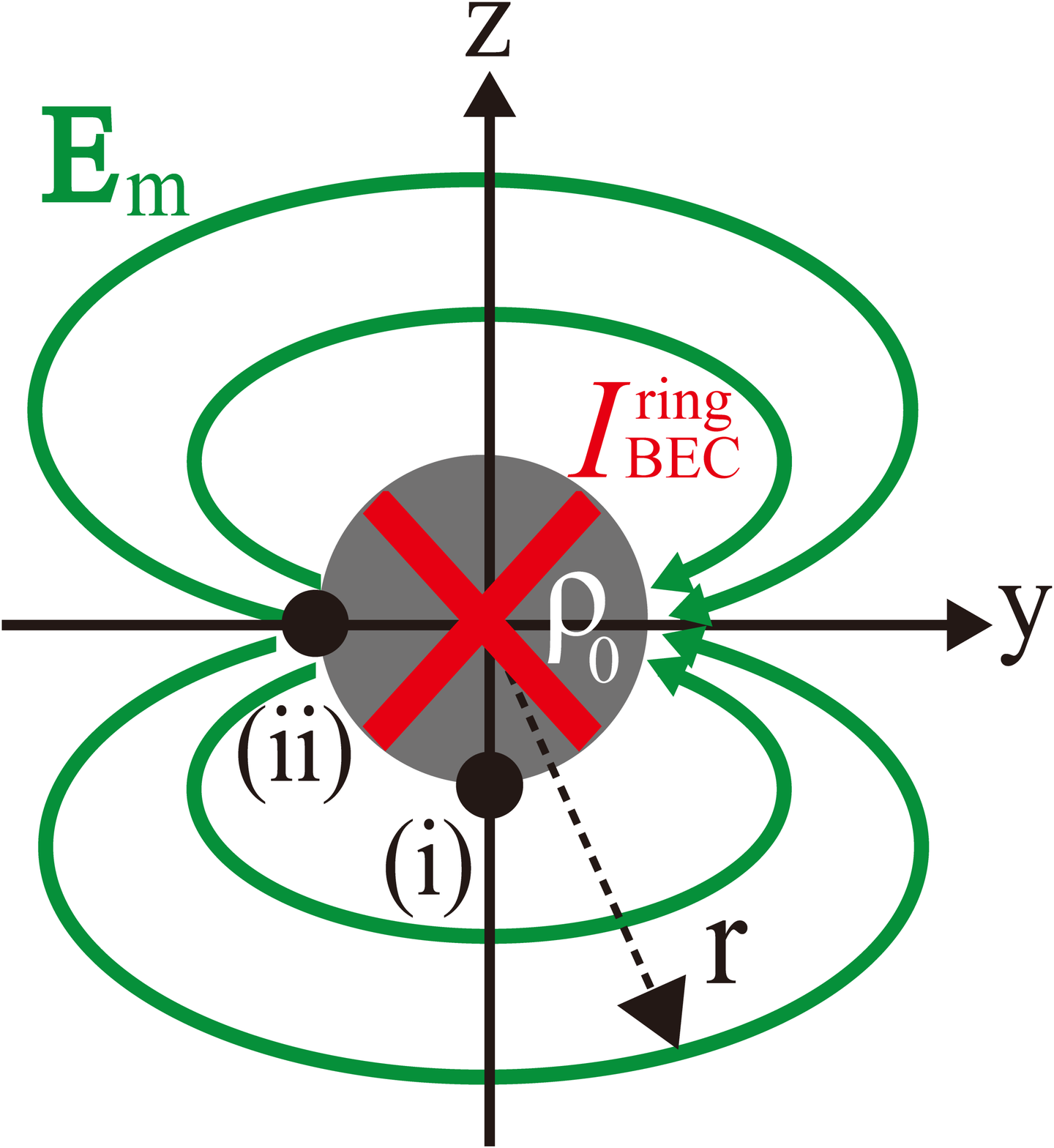}
\caption{(Color online)
A schematic picture of the cross-section of the cylindrical wire.
The electric dipole field $  {\bf{E}}_{\rm{m}}  $ arises from the persistent magnon-BEC current  $   I_{\rm{BEC}}^{\rm{ring}} $ (\textit{i.e.} the magnetic dipole  steady current) flowing in the magnon-BEC ring. 
This leads to a measurable voltage drop  $ V_{\rm{m}}   $ between the points (i) and (ii).
The points are (i) $ (y, z)= (0, -r_0) $ and (ii) $ (y, z)= (-r_0, 0) $.
 \label{fig:dipole} }
\end{center}
\end{figure}

\subsection{Magnon-BEC ring}
\label{subsec:BECring}

Next, we apply this result to the magnon-BEC ring [Fig.  \ref{fig:SQUID} (a)]. Assuming single-valuedness of the BEC wave function around the ring, in analogy to
superconducting rings, 
a standard analysis gives [see also Eq. (\ref{eqn:acac})]
\begin{subequations}
\begin{eqnarray}
\theta _{\rm{A{\mathchar`-}C}}
&= &
 \frac{g \mu_{\rm{B}} }{\hbar c^2} \oint \textrm{d} {\bf l} \cdot \left( {\bf E} \times {\bf e}_z\right)   \\
&= & 2\pi \frac{\phi}{\phi_0},
\label{eq:AC_SQUID}
\end{eqnarray}
\end{subequations}
where `electric-gradient flux' through the ring is defined as 
\begin{eqnarray}
 {\phi}\equiv \int \textrm{d} A {\bf e}_z \cdot \nabla \times ({\bf E} \times {\bf e}_z),  
 \label{eq:e-gflux}
\end{eqnarray}
and the `electric flux quantum'\cite{wei2}  $\phi_0 \equiv  hc^2/(g\mu_{\rm{B}})$. 
The electric-gradient flux ${\phi}  $ is quantized as $  {\phi}  = p \phi_0 $, where the integer $p$ is the phase winding number of the closed path around the magnon-BEC ring. When the quantization condition is not satisfied, the current that flows in the ring is not steady. However, these non-steady variations of the current away from its equilibrium value are small, on the relative order of $1/p \ll 1$ (see estimates below).
This is in contrast to a superconducting ring where the quantization of the magnetic flux
is achieved by the magnetic field of the supercurrent itself to compensate for  variations on the external magnetic flux.~\cite{Zurich} 

We note that for ${\bf E}({\bf r}) = {\bf E}(x,y)$, the expression for $\phi$ simplifies to $\phi = -  \int \textrm{d} A \nabla\cdot {\bf E}$. For an electric field of the form ${\bf E}(\rho , \varphi ) =  {\cal{E}}  \rho  {\bf e}_{\rho }$ [Fig.  \ref{fig:SQUID} (a)], we find that $\phi = - \pi R^2   {\cal{E}}  $ for a ring of radius $R$.
We now suppose that the ring consists of the cylindrical wire whose cross-section is $ \pi  \rho_0 ^2$ with the radius $\rho_0 $ [Fig.  \ref{fig:SQUID} (b)]. 
Consequently, using Eq. (\ref{eqn:BECcurrent}), $\theta _{\rm{A{\mathchar`-}C}}^{\rm{chain}}=[\alpha /(\pi R)] \theta _{\rm{A{\mathchar`-}C}} $, and $N_{\rm{BEC}} = [(\pi  \rho_0^2)\alpha] n_{\rm{BEC}}$,
the magnitude of the persistent magnon-BEC current in the magnon-BEC ring becomes  
$\mid   I_{\rm{BEC}}^{\rm{ring}}  \mid  =   2 {\pi}  g \mu _{\rm{B}}   \mid J/\hbar \mid   S  \rho_0^2 \alpha  n_{\rm{BEC}} \mid  {\rm{sin}}[2 \alpha   \phi /( R  \phi_0)] \mid  $, where $n_{\rm{BEC}}$ is the density of condensed magnons.\cite{demokritov}

\begin{figure}[h]
\begin{center}
\includegraphics[width=7cm,clip]{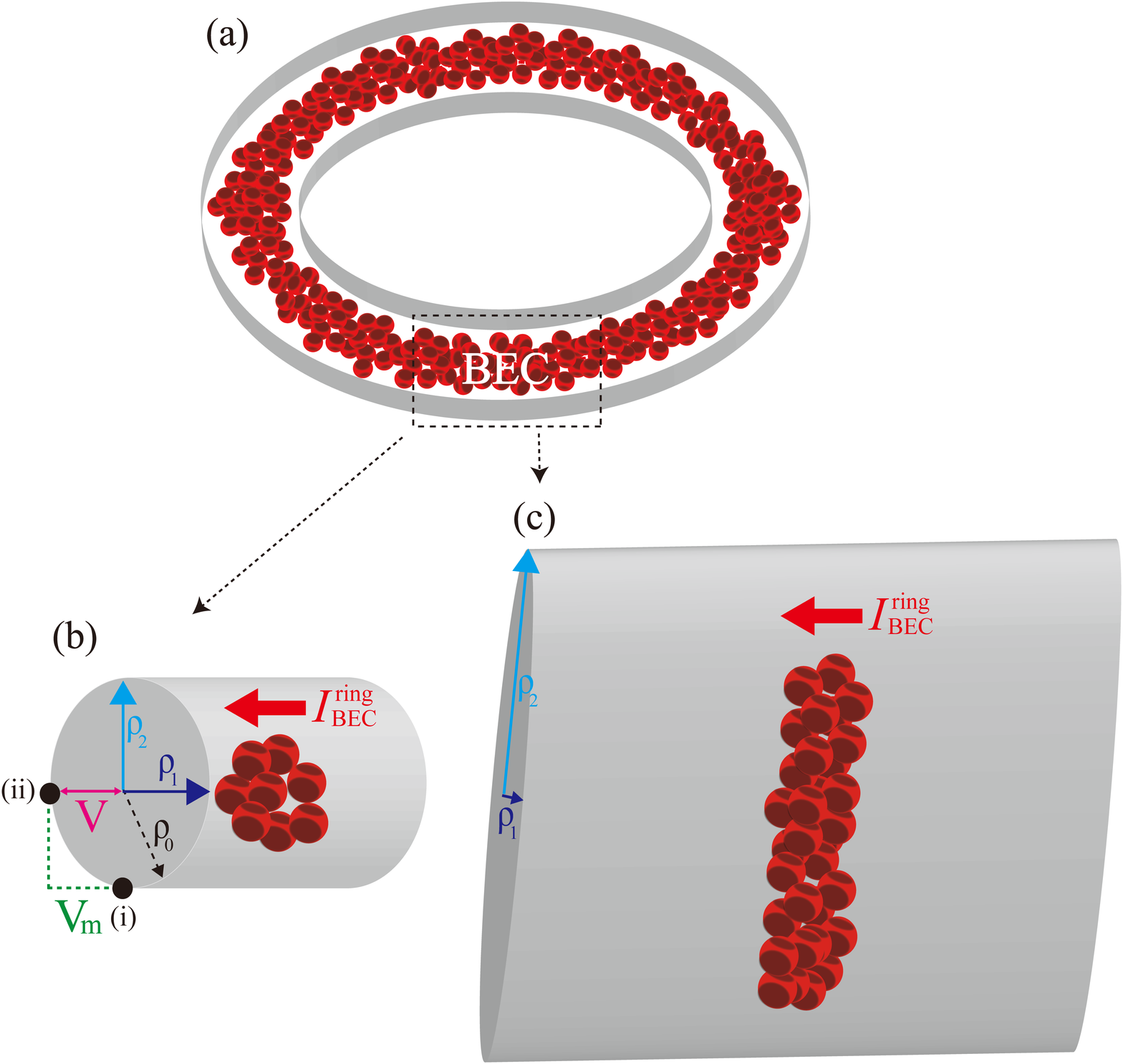}
\caption{(Color online)
(a) A schematic picture of a magnon-BEC ring.
(b) An enlarged view of the isotropic cylindrical wire that forms the magnon-BEC ring ($ r_0  = \rho _0 = \rho _1  =  \rho _2  $). The cross-section is $  \pi  \rho _0^2 $. The voltage drop due to the persistent magnon-BEC current $ I_{\rm{BEC}}^{\rm{ring}}$ is $V_{\rm{m}} $ and the one arising directly from the applied electric field is $V$. The points are (i) $ (y, z)= (0, -r_0) $ and (ii) $ (y, z)= (-r_0, 0) $  (see also Fig. \ref{fig:dipole}). 
(c)  An enlarged view of the anisotropic cylindrical wire that forms the ring ($  \rho _1  \ll  \rho _2  $; {\textit{e.g.}} $ \rho _2  = 100 \rho _1  $). Consequently, the voltage drop $V$ directly due to the applied electric field becomes much smaller (about $10^{-2}$ times) than the one of (b). The cross-section remains about the same with (b). Therefore it generates about the same amount of the persistent magnon-BEC current $ I_{\rm{BEC}}^{\rm{ring}}$ as for that of (b).
\label{fig:RevisedSQUID} }
\end{center}
\end{figure}

\subsection{Experimental detection}
\label{subsec:detection}

Finally, we show how to directly measure the persistent magnon-BEC current in the ring. 
Note that the persistent magnon-BEC current is a steady flow of  magnetic dipoles (\textit{i.e.} magnons with  magnetic moment $g\mu_{\rm{B}}  {\bf e}_z$) and the moving magnetic dipoles produce electric dipole fields.\cite{dipole} The magnitude of the resulting electric dipole field in Fig.  \ref{fig:dipole}  is\cite{magnon2} $ |{\bf{E}}_{\rm{m}} (r)  |= \mu _0 \mid   I_{\rm{BEC}}^{\rm{ring}}   \mid / (2 \pi  r^2 )  $.
Assuming\cite{demokritov} $\rho_0 = 1$ mm, $ r_0 = 1 $ mm, $ R \approx  10 $ mm, $ \alpha =1 $ \AA, $ n_{\rm{BEC}}= 10^{19} $ cm$^{-3}$, 
$   {\cal{E}} R   \approx  5   $ GV/m, $ g = 2$, $ S= 2$, and  $ \mid J \mid = 0.1 $ eV, 
the phase winding number becomes $p \approx 50 $ and  an electric field $ \mid  {\bf{E}}_{\rm{m}}  (r_0)   \mid  \sim 4 \mu  $V/m  is generated. This leads to a voltage drop $ V_{\rm{m}}   \sim  4   $nV between the points (i) and (ii) in Fig.  \ref{fig:dipole}. 
Although being small, such a value is within experimental reach and  due to the macroscopic coherence of magnon-BECs,
the value is actually much larger than the one which has been predicted  in spin chain systems ($\textit{i.e.}$ non-condensed  magnonic systems).\cite{magnon2} 

Regarding the experimental detection of the persistent magnon-BEC current, one concern\cite{Zurich}
 might be that the ratio ${\cal{R}} \equiv  \mid  V_{\rm{m}}/V  \mid $ would be too small, where $V_{\rm{m}} $  is the voltage drop due to the persistent magnon-BEC current and $V$  is the one arising directly from the applied electric field. In the case of Fig. \ref{fig:RevisedSQUID} (b) [or Fig. \ref{fig:SQUID} (b)], it amounts to ${\cal{R}}_{\rm{b}} \approx  1$ nV/ $1$ MV.
We remark, however, that this ratio could be improved by using an anisotropic cylindrical wire shown in Fig. \ref{fig:RevisedSQUID} (c), where $\rho _1 \ll  \rho _2 $ ({\textit{e.g.}} $  \rho _2 = 100 \rho _1$) and the cross-sectional area remains about the same as for Fig. \ref{fig:RevisedSQUID} (b). Therefore, this cylinder geometry generates about the same amount of  persistent magnon-BEC current $ I_{\rm{BEC}}^{\rm{ring}}$ as that of Fig. \ref{fig:RevisedSQUID} (b) with substantially reduced magnitude $V$, while, simultaneously,  $V_{\rm{m}} $ is  increased.
Consequently, the ratio ${\cal{R}}_{\rm{c}}$ is also much improved in  Fig. \ref{fig:RevisedSQUID} (c). Indeed, a rough estimate gives ${\cal{R}}_{\rm{c}}    / {\cal{R}}_{\rm{b}} \approx  10^4 $. 

Lastly, we remark that the above method for the detection of magnon-BEC currents via the voltage drop is applicable also to the Josephson junction shown in Fig. \ref{fig:Josephson}. The Josephson magnon-BEC currents are generated at the interface and flow between the FIs. Consequently, a voltage drop arises at the interface. When an ac or dc Josephson effect is generated, the resulting voltage drop becomes also ac or dc. 
Thus, the experimental detection of such ac and dc effects would be in principle possible.
As discussed in Sec. \ref{sec:weak}, the A-C phase is essential for the generation of the dc Josephson effect [Fig. \ref{fig:DCbreakdownACphase} (b)] as well as for the persistent  current in the ring. Therefore, the detection of the dc Josephson effect or the persistent magnon-BEC current through the resulting dc voltage drop would both be a manifestation of the A-C phase. 
When MQST occurs, the oscillation ({\textit{i.e.}} period) of the ac effect increases as shown in Fig. \ref{fig:MQST} (d) [see also Fig. \ref{fig:MQST} (a) for the usual ac effect]. 
Thus, all these phenomena ({\textit{i.e.}} ac or dc Josephson effects and MQST) are in principle experimentally detectable via the Josephson effect.
Indeed, to estimate the voltage we assume the junction formed by a cylindrical wire as shown in Fig. \ref{fig:SQUID} (b), with parameter values~\cite{demokritov} $\rho_0 = 1$ mm, $ r_0 = 1 $ mm, $\Delta x \approx 10 $\AA, $n_{\rm{T}} \sim  n_{\rm{BEC}}= 10^{19} $ cm$^{-3}$, $ g = 2$, $ S= 2$, and  $  J_{\rm{ex}} = 0.25 \mu $eV. 
Then, the corresponding  drop  of the voltage amplitude amounts to $0 \leq  V_{\rm{m}} \leq  1 \mu $V at the interface generated  by the ac or dc Josephson effects or by the MQST.

Given these estimates, we conclude that the experimental detection of  magnon-BEC currents, while being challenging, seems within reach.

\section{Summary}
\label{sec:summary}
Starting from the microscopic Heisenberg Hamiltonian, we have derived the equations of motion that describe the tunneling of a Josephson magnon current through a junction connecting two quasi-equilibrium magnon-BECs. We have discussed the occurrence of the ac Josephson effect and macroscopic quantum self-trapping in such junctions. We have proposed a method to generate a dc Josephson magnon current using time-dependent magnetic fields, and have shown that the application of an electric field to the system influence the physics of the junction through the Aharonov-Casher phase. 
Finally, we have introduced a magnon-BEC ring and have proposed a method (based on induced voltage differences) to directly measure persistent magnon-BEC currents in the ring due to the Aharonov-Casher phase. Likewise, the same method can be used to detect the Josephson effects.

\begin{acknowledgments}
We would like to thank  Yaroslav Tserkovnyak for fruitful discussions and Andreas Schilling for useful comments on our manuscript.
We acknowledge support   by the Swiss NSF, the NCCR QSIT ETHZ-Basel, and the FP7-ICT project  "ELFOS" (K. N., K. A. v. H., and D. L.), and from the ANR under Contract No. DYMESYS (ANR 2011-IS04-001-01) (P.S.). One of the authors (K. N.) is grateful to the Grant-in-Aid for JSPS Research Fellow (No. 25-2747) and the young researchers exchange program supported by the Yukawa Institute for Theoretical Physics. 
\end{acknowledgments}

\bibliography{PumpingRef}

\begin{thebibliography}{49}
\expandafter\ifx\csname natexlab\endcsname\relax\def\natexlab#1{#1}\fi
\expandafter\ifx\csname bibnamefont\endcsname\relax
  \def\bibnamefont#1{#1}\fi
\expandafter\ifx\csname bibfnamefont\endcsname\relax
  \def\bibfnamefont#1{#1}\fi
\expandafter\ifx\csname citenamefont\endcsname\relax
  \def\citenamefont#1{#1}\fi
\expandafter\ifx\csname url\endcsname\relax
  \def\url#1{\texttt{#1}}\fi
\expandafter\ifx\csname urlprefix\endcsname\relax\def\urlprefix{URL }\fi
\providecommand{\bibinfo}[2]{#2}
\providecommand{\eprint}[2][]{\url{#2}}

\bibitem[{\citenamefont{Awschalom and
  Flatt$\acute{\text{e}}$}(2007)}]{awschalom}
\bibinfo{author}{\bibfnamefont{D.~D.} \bibnamefont{Awschalom}}
  \bibnamefont{and} \bibinfo{author}{\bibfnamefont{M.~E.}
  \bibnamefont{Flatt$\acute{\text{e}}$}}, \bibinfo{journal}{Nat. Phys.}
  \textbf{\bibinfo{volume}{3}}, \bibinfo{pages}{153} (\bibinfo{year}{2007}).

\bibitem[{\citenamefont{Zutic et~al.}(2004)\citenamefont{Zutic, Fabian, and
  Sarma}}]{mod}
\bibinfo{author}{\bibfnamefont{I.}~\bibnamefont{Zutic}},
  \bibinfo{author}{\bibfnamefont{J.}~\bibnamefont{Fabian}}, \bibnamefont{and}
  \bibinfo{author}{\bibfnamefont{S.~D.} \bibnamefont{Sarma}},
  \bibinfo{journal}{Rev. Mod. Phys.} \textbf{\bibinfo{volume}{76}},
  \bibinfo{pages}{323} (\bibinfo{year}{2004}).

\bibitem[{\citenamefont{Meier and Loss}(2003)}]{magnon2}
\bibinfo{author}{\bibfnamefont{F.}~\bibnamefont{Meier}} \bibnamefont{and}
  \bibinfo{author}{\bibfnamefont{D.}~\bibnamefont{Loss}},
  \bibinfo{journal}{Phys. Rev. Lett.} \textbf{\bibinfo{volume}{90}},
  \bibinfo{pages}{167204} (\bibinfo{year}{2003}).

\bibitem[{\citenamefont{Silsbee et~al.}(1979)\citenamefont{Silsbee, Janossy,
  and Monod}}]{silsbee}
\bibinfo{author}{\bibfnamefont{R.~H.} \bibnamefont{Silsbee}},
  \bibinfo{author}{\bibfnamefont{A.}~\bibnamefont{Janossy}}, \bibnamefont{and}
  \bibinfo{author}{\bibfnamefont{P.}~\bibnamefont{Monod}},
  \bibinfo{journal}{Phys. Rev. B} \textbf{\bibinfo{volume}{19}},
  \bibinfo{pages}{4382} (\bibinfo{year}{1979}).

\bibitem[{\citenamefont{Saitoh et~al.}(2006)\citenamefont{Saitoh, Ueda,
  Miyajima, and Tatara}}]{ISHE1}
\bibinfo{author}{\bibfnamefont{E.}~\bibnamefont{Saitoh}},
  \bibinfo{author}{\bibfnamefont{M.}~\bibnamefont{Ueda}},
  \bibinfo{author}{\bibfnamefont{H.}~\bibnamefont{Miyajima}}, \bibnamefont{and}
  \bibinfo{author}{\bibfnamefont{G.}~\bibnamefont{Tatara}},
  \bibinfo{journal}{Appl. Phys. Lett.} \textbf{\bibinfo{volume}{88}},
  \bibinfo{pages}{182509} (\bibinfo{year}{2006}).

\bibitem[{\citenamefont{Brataas et~al.}(2002)\citenamefont{Brataas,
  Tserkovnyak, Bauer, and Halperin}}]{battery}
\bibinfo{author}{\bibfnamefont{A.}~\bibnamefont{Brataas}},
  \bibinfo{author}{\bibfnamefont{Y.}~\bibnamefont{Tserkovnyak}},
  \bibinfo{author}{\bibfnamefont{G.~E.~W.} \bibnamefont{Bauer}},
  \bibnamefont{and} \bibinfo{author}{\bibfnamefont{B.~I.}
  \bibnamefont{Halperin}}, \bibinfo{journal}{Phys. Rev. B}
  \textbf{\bibinfo{volume}{66}}, \bibinfo{pages}{060404(R)}
  (\bibinfo{year}{2002}).

\bibitem[{\citenamefont{Tserkovnyak et~al.}(2005)\citenamefont{Tserkovnyak,
  Brataas, Bauer, and Halperin}}]{mod2}
\bibinfo{author}{\bibfnamefont{Y.}~\bibnamefont{Tserkovnyak}},
  \bibinfo{author}{\bibfnamefont{A.}~\bibnamefont{Brataas}},
  \bibinfo{author}{\bibfnamefont{G.~E.~W.} \bibnamefont{Bauer}},
  \bibnamefont{and} \bibinfo{author}{\bibfnamefont{B.~I.}
  \bibnamefont{Halperin}}, \bibinfo{journal}{Rev. Mod. Phys.}
  \textbf{\bibinfo{volume}{77}}, \bibinfo{pages}{1375} (\bibinfo{year}{2005}).

\bibitem[{\citenamefont{Takeuchi et~al.}(2010)\citenamefont{Takeuchi, Hosono,
  and Tatara}}]{takeuchi}
\bibinfo{author}{\bibfnamefont{A.}~\bibnamefont{Takeuchi}},
  \bibinfo{author}{\bibfnamefont{K.}~\bibnamefont{Hosono}}, \bibnamefont{and}
  \bibinfo{author}{\bibfnamefont{G.}~\bibnamefont{Tatara}},
  \bibinfo{journal}{Phys. Rev. B} \textbf{\bibinfo{volume}{81}},
  \bibinfo{pages}{144405} (\bibinfo{year}{2010}).

\bibitem[{\citenamefont{Nakata and Tatara}(2011)}]{nakatatatara}
\bibinfo{author}{\bibfnamefont{K.}~\bibnamefont{Nakata}} \bibnamefont{and}
  \bibinfo{author}{\bibfnamefont{G.}~\bibnamefont{Tatara}},
  \bibinfo{journal}{J. Phys. Soc. Jpn.} \textbf{\bibinfo{volume}{80}},
  \bibinfo{pages}{054602} (\bibinfo{year}{2011}).

\bibitem[{\citenamefont{Nakata}(2012)}]{QSP}
\bibinfo{author}{\bibfnamefont{K.}~\bibnamefont{Nakata}}, \bibinfo{journal}{J.
  Phys. Soc. Jpn.} \textbf{\bibinfo{volume}{81}}, \bibinfo{pages}{064717}
  (\bibinfo{year}{2012}).

\bibitem[{\citenamefont{Nakata}(2013)}]{QSPtemp}
\bibinfo{author}{\bibfnamefont{K.}~\bibnamefont{Nakata}}, \bibinfo{journal}{J.
  Phys.: Condens. Matter.} \textbf{\bibinfo{volume}{25}},
  \bibinfo{pages}{116005} (\bibinfo{year}{2013}).

\bibitem[{\citenamefont{Kajiwara et~al.}(2010)\citenamefont{Kajiwara, Harii,
  Takahashi, Ohe, Uchida, Mizuguchi, Umezawa, Kawai, Ando, Takanashi
  et~al.}}]{spinwave}
\bibinfo{author}{\bibfnamefont{Y.}~\bibnamefont{Kajiwara}},
  \bibinfo{author}{\bibfnamefont{K.}~\bibnamefont{Harii}},
  \bibinfo{author}{\bibfnamefont{S.}~\bibnamefont{Takahashi}},
  \bibinfo{author}{\bibfnamefont{J.}~\bibnamefont{Ohe}},
  \bibinfo{author}{\bibfnamefont{K.}~\bibnamefont{Uchida}},
  \bibinfo{author}{\bibfnamefont{M.}~\bibnamefont{Mizuguchi}},
  \bibinfo{author}{\bibfnamefont{H.}~\bibnamefont{Umezawa}},
  \bibinfo{author}{\bibfnamefont{H.}~\bibnamefont{Kawai}},
  \bibinfo{author}{\bibfnamefont{K.}~\bibnamefont{Ando}},
  \bibinfo{author}{\bibfnamefont{K.}~\bibnamefont{Takanashi}},
  \bibnamefont{et~al.}, \bibinfo{journal}{Nature}
  \textbf{\bibinfo{volume}{464}}, \bibinfo{pages}{262} (\bibinfo{year}{2010}).

\bibitem[{\citenamefont{Tserkovnyak}(2013)}]{TserkovnyakNatNano}
\bibinfo{author}{\bibfnamefont{Y.}~\bibnamefont{Tserkovnyak}},
  \bibinfo{journal}{Nat. Nanotechnology} \textbf{\bibinfo{volume}{8}},
  \bibinfo{pages}{706} (\bibinfo{year}{2013}).

\bibitem[{\citenamefont{Bender et~al.}(2012)\citenamefont{Bender, Duine, and
  Tserkovnyak}}]{bender}
\bibinfo{author}{\bibfnamefont{S.~A.} \bibnamefont{Bender}},
  \bibinfo{author}{\bibfnamefont{R.~A.} \bibnamefont{Duine}}, \bibnamefont{and}
  \bibinfo{author}{\bibfnamefont{Y.}~\bibnamefont{Tserkovnyak}},
  \bibinfo{journal}{Phys. Rev. Lett.} \textbf{\bibinfo{volume}{108}},
  \bibinfo{pages}{246601} (\bibinfo{year}{2012}).

\bibitem[{\citenamefont{Takei and Tserkovnyak}(2014)}]{Takei}
\bibinfo{author}{\bibfnamefont{S.}~\bibnamefont{Takei}} \bibnamefont{and}
  \bibinfo{author}{\bibfnamefont{Y.}~\bibnamefont{Tserkovnyak}},
  \bibinfo{journal}{Phys. Rev. Lett.} \textbf{\bibinfo{volume}{112}},
  \bibinfo{pages}{227201} (\bibinfo{year}{2014}).

\bibitem[{\citenamefont{Chen et~al.}(2014)\citenamefont{Chen, Horsch, and
  Manske}}]{wei}
\bibinfo{author}{\bibfnamefont{W.}~\bibnamefont{Chen}},
  \bibinfo{author}{\bibfnamefont{P.}~\bibnamefont{Horsch}}, \bibnamefont{and}
  \bibinfo{author}{\bibfnamefont{D.}~\bibnamefont{Manske}},
  \bibinfo{journal}{Phys. Rev. B} \textbf{\bibinfo{volume}{89}},
  \bibinfo{pages}{064427} (\bibinfo{year}{2014}).

\bibitem[{\citenamefont{Trauzettel et~al.}(2008)\citenamefont{Trauzettel,
  Simon, and Loss}}]{Trauzettel}
\bibinfo{author}{\bibfnamefont{B.}~\bibnamefont{Trauzettel}},
  \bibinfo{author}{\bibfnamefont{P.}~\bibnamefont{Simon}}, \bibnamefont{and}
  \bibinfo{author}{\bibfnamefont{D.}~\bibnamefont{Loss}},
  \bibinfo{journal}{Phys. Rev. Lett.} \textbf{\bibinfo{volume}{101}},
  \bibinfo{pages}{017202} (\bibinfo{year}{2008}).

\bibitem[{\citenamefont{Zapf et~al.}(2014)\citenamefont{Zapf, Jaime, and
  Batista}}]{BatistaBEC}
\bibinfo{author}{\bibfnamefont{V.}~\bibnamefont{Zapf}},
  \bibinfo{author}{\bibfnamefont{M.}~\bibnamefont{Jaime}}, \bibnamefont{and}
  \bibinfo{author}{\bibfnamefont{C.~D.} \bibnamefont{Batista}},
  \bibinfo{journal}{Rev. Mod. Phys.} \textbf{\bibinfo{volume}{86}},
  \bibinfo{pages}{563} (\bibinfo{year}{2014}).

\bibitem[{\citenamefont{Demokritov et~al.}(2006)\citenamefont{Demokritov,
  Demidov, Dzyapko, Melkov, Serga, Hillebrands, and Slavin}}]{demokritov}
\bibinfo{author}{\bibfnamefont{S.~O.} \bibnamefont{Demokritov}},
  \bibinfo{author}{\bibfnamefont{V.~E.} \bibnamefont{Demidov}},
  \bibinfo{author}{\bibfnamefont{O.}~\bibnamefont{Dzyapko}},
  \bibinfo{author}{\bibfnamefont{G.~A.} \bibnamefont{Melkov}},
  \bibinfo{author}{\bibfnamefont{A.~A.} \bibnamefont{Serga}},
  \bibinfo{author}{\bibfnamefont{B.}~\bibnamefont{Hillebrands}},
  \bibnamefont{and} \bibinfo{author}{\bibfnamefont{A.~N.}
  \bibnamefont{Slavin}}, \bibinfo{journal}{Nature}
  \textbf{\bibinfo{volume}{443}}, \bibinfo{pages}{430} (\bibinfo{year}{2006}).

\bibitem[{\citenamefont{Josephson}(1962)}]{Josephson}
\bibinfo{author}{\bibfnamefont{B.~D.} \bibnamefont{Josephson}},
  \bibinfo{journal}{Phys. Lett.} \textbf{\bibinfo{volume}{1}},
  \bibinfo{pages}{251} (\bibinfo{year}{1962}).

\bibitem[{\citenamefont{Tserkovnyak and Loss}(2011)}]{ytDL}
\bibinfo{author}{\bibfnamefont{Y.}~\bibnamefont{Tserkovnyak}} \bibnamefont{and}
  \bibinfo{author}{\bibfnamefont{D.}~\bibnamefont{Loss}},
  \bibinfo{journal}{Phys. Rev. A} \textbf{\bibinfo{volume}{84}},
  \bibinfo{pages}{032333} (\bibinfo{year}{2011}).

\bibitem[{\citenamefont{Tatara and Garcia}(2003)}]{TataraJosephson}
\bibinfo{author}{\bibfnamefont{G.}~\bibnamefont{Tatara}} \bibnamefont{and}
  \bibinfo{author}{\bibfnamefont{N.}~\bibnamefont{Garcia}},
  \bibinfo{journal}{Phys. Rev. Lett.} \textbf{\bibinfo{volume}{91}},
  \bibinfo{pages}{076806} (\bibinfo{year}{2003}).

\bibitem[{\citenamefont{Schilling and Grundmann}(2012)}]{Schilling}
\bibinfo{author}{\bibfnamefont{A.}~\bibnamefont{Schilling}} \bibnamefont{and}
  \bibinfo{author}{\bibfnamefont{H.}~\bibnamefont{Grundmann}},
  \bibinfo{journal}{Ann. Phys.} \textbf{\bibinfo{volume}{327}},
  \bibinfo{pages}{2301} (\bibinfo{year}{2012}).

\bibitem[{\citenamefont{Troncoso and $\acute{\rm{A}}$.
  S.~N$\acute{\rm{u}}$$\tilde{\rm{n}}$ez}()}]{troncoso2}
\bibinfo{author}{\bibfnamefont{R.~E.} \bibnamefont{Troncoso}} \bibnamefont{and}
  \bibinfo{author}{\bibnamefont{$\acute{\rm{A}}$.
  S.~N$\acute{\rm{u}}$$\tilde{\rm{n}}$ez}}, \bibinfo{note}{arXiv:1305.4285}.

\bibitem[{\citenamefont{Aharonov and Casher}(1984)}]{casher}
\bibinfo{author}{\bibfnamefont{Y.}~\bibnamefont{Aharonov}} \bibnamefont{and}
  \bibinfo{author}{\bibfnamefont{A.}~\bibnamefont{Casher}},
  \bibinfo{journal}{Phys. Rev. Lett.} \textbf{\bibinfo{volume}{53}},
  \bibinfo{pages}{319} (\bibinfo{year}{1984}).

\bibitem[{\citenamefont{Loss et~al.}(1990)\citenamefont{Loss, Goldbart, and
  Balatsky}}]{LossPersistent}
\bibinfo{author}{\bibfnamefont{D.}~\bibnamefont{Loss}},
  \bibinfo{author}{\bibfnamefont{P.}~\bibnamefont{Goldbart}}, \bibnamefont{and}
  \bibinfo{author}{\bibfnamefont{A.~V.} \bibnamefont{Balatsky}},
  \bibinfo{journal}{Phys. Rev. Lett.} \textbf{\bibinfo{volume}{65}},
  \bibinfo{pages}{1655} (\bibinfo{year}{1990}).

\bibitem[{\citenamefont{Loss and Goldbart}(1992)}]{LossPersistent2}
\bibinfo{author}{\bibfnamefont{D.}~\bibnamefont{Loss}} \bibnamefont{and}
  \bibinfo{author}{\bibfnamefont{P.~M.} \bibnamefont{Goldbart}},
  \bibinfo{journal}{Phys. Rev. B} \textbf{\bibinfo{volume}{45}},
  \bibinfo{pages}{13544} (\bibinfo{year}{1992}).

\bibitem[{\citenamefont{Loss and Goldbart}(1996)}]{dipole}
\bibinfo{author}{\bibfnamefont{D.}~\bibnamefont{Loss}} \bibnamefont{and}
  \bibinfo{author}{\bibfnamefont{P.~M.} \bibnamefont{Goldbart}},
  \bibinfo{journal}{Phys. Lett. A} \textbf{\bibinfo{volume}{215}},
  \bibinfo{pages}{197} (\bibinfo{year}{1996}).

\bibitem[{\citenamefont{Holstein and Primakoff}(1940)}]{HP}
\bibinfo{author}{\bibfnamefont{T.}~\bibnamefont{Holstein}} \bibnamefont{and}
  \bibinfo{author}{\bibfnamefont{H.}~\bibnamefont{Primakoff}},
  \bibinfo{journal}{Phys. Rev.} \textbf{\bibinfo{volume}{58}},
  \bibinfo{pages}{1098} (\bibinfo{year}{1940}).

\bibitem[{\citenamefont{Zapata et~al.}(1998)\citenamefont{Zapata, Sols, and
  Leggett}}]{Zapata}
\bibinfo{author}{\bibfnamefont{I.}~\bibnamefont{Zapata}},
  \bibinfo{author}{\bibfnamefont{F.}~\bibnamefont{Sols}}, \bibnamefont{and}
  \bibinfo{author}{\bibfnamefont{A.~J.} \bibnamefont{Leggett}},
  \bibinfo{journal}{Phys. Rev. A} \textbf{\bibinfo{volume}{57}},
  \bibinfo{pages}{R28} (\bibinfo{year}{1998}).

\bibitem[{\citenamefont{Aharonov and Bohm}(1959)}]{bohm}
\bibinfo{author}{\bibfnamefont{Y.}~\bibnamefont{Aharonov}} \bibnamefont{and}
  \bibinfo{author}{\bibfnamefont{D.}~\bibnamefont{Bohm}},
  \bibinfo{journal}{Phys. Rev.} \textbf{\bibinfo{volume}{115}},
  \bibinfo{pages}{485} (\bibinfo{year}{1959}).

\bibitem[{\citenamefont{Mignani}(1991, and references therein)}]{Mignani}
\bibinfo{author}{\bibfnamefont{R.}~\bibnamefont{Mignani}}, \bibinfo{journal}{J.
  Phys. A: Math. Gen.} \textbf{\bibinfo{volume}{24}}, \bibinfo{pages}{L421}
  (\bibinfo{year}{1991, and references therein}).

\bibitem[{\citenamefont{Hea and McKellarb}(1991)}]{Hea}
\bibinfo{author}{\bibfnamefont{X.-G.} \bibnamefont{Hea}} \bibnamefont{and}
  \bibinfo{author}{\bibfnamefont{B.}~\bibnamefont{McKellarb}},
  \bibinfo{journal}{Phys. Lett. B} \textbf{\bibinfo{volume}{264}},
  \bibinfo{pages}{129} (\bibinfo{year}{1991}).

\bibitem[{\citenamefont{Smerzi et~al.}(1997)\citenamefont{Smerzi, Fantoni,
  Giovanazzi, and Shenoy}}]{smerzi}
\bibinfo{author}{\bibfnamefont{A.}~\bibnamefont{Smerzi}},
  \bibinfo{author}{\bibfnamefont{S.}~\bibnamefont{Fantoni}},
  \bibinfo{author}{\bibfnamefont{S.}~\bibnamefont{Giovanazzi}},
  \bibnamefont{and} \bibinfo{author}{\bibfnamefont{S.~R.}
  \bibnamefont{Shenoy}}, \bibinfo{journal}{Phys. Rev. Lett.}
  \textbf{\bibinfo{volume}{79}}, \bibinfo{pages}{4950} (\bibinfo{year}{1997}).

\bibitem[{\citenamefont{Levy et~al.}(2007)\citenamefont{Levy, Lahoud, Shomroni,
  and Steinhauer}}]{Levy}
\bibinfo{author}{\bibfnamefont{S.}~\bibnamefont{Levy}},
  \bibinfo{author}{\bibfnamefont{E.}~\bibnamefont{Lahoud}},
  \bibinfo{author}{\bibfnamefont{I.}~\bibnamefont{Shomroni}}, \bibnamefont{and}
  \bibinfo{author}{\bibfnamefont{J.}~\bibnamefont{Steinhauer}},
  \bibinfo{journal}{Nature} \textbf{\bibinfo{volume}{449}},
  \bibinfo{pages}{579} (\bibinfo{year}{2007}).

\bibitem[{\citenamefont{Raghavan et~al.}(1999)\citenamefont{Raghavan, Smerzi,
  Fantoni, and Shenoy}}]{smerzi2}
\bibinfo{author}{\bibfnamefont{S.}~\bibnamefont{Raghavan}},
  \bibinfo{author}{\bibfnamefont{A.}~\bibnamefont{Smerzi}},
  \bibinfo{author}{\bibfnamefont{S.}~\bibnamefont{Fantoni}}, \bibnamefont{and}
  \bibinfo{author}{\bibfnamefont{S.~R.} \bibnamefont{Shenoy}},
  \bibinfo{journal}{Phys. Rev. A} \textbf{\bibinfo{volume}{59}},
  \bibinfo{pages}{620} (\bibinfo{year}{1999}).

\bibitem[{\citenamefont{Giovanazzi et~al.}(2000)\citenamefont{Giovanazzi,
  Smerzi, and Fantoni}}]{smerzi3}
\bibinfo{author}{\bibfnamefont{S.}~\bibnamefont{Giovanazzi}},
  \bibinfo{author}{\bibfnamefont{A.}~\bibnamefont{Smerzi}}, \bibnamefont{and}
  \bibinfo{author}{\bibfnamefont{S.}~\bibnamefont{Fantoni}},
  \bibinfo{journal}{Phys. Rev. Lett.} \textbf{\bibinfo{volume}{84}},
  \bibinfo{pages}{4521} (\bibinfo{year}{2000}).

\bibitem[{\citenamefont{Albiez et~al.}(2005)\citenamefont{Albiez, Gati,
  F${\ddot{{\rm{o}}} }$lling, Hunsmann, Cristiani, and Oberthaler}}]{albiez}
\bibinfo{author}{\bibfnamefont{M.}~\bibnamefont{Albiez}},
  \bibinfo{author}{\bibfnamefont{R.}~\bibnamefont{Gati}},
  \bibinfo{author}{\bibfnamefont{J.}~\bibnamefont{F${\ddot{{\rm{o}}} }$lling}},
  \bibinfo{author}{\bibfnamefont{S.}~\bibnamefont{Hunsmann}},
  \bibinfo{author}{\bibfnamefont{M.}~\bibnamefont{Cristiani}},
  \bibnamefont{and} \bibinfo{author}{\bibfnamefont{M.~K.}
  \bibnamefont{Oberthaler}}, \bibinfo{journal}{Phys. Rev. Lett.}
  \textbf{\bibinfo{volume}{95}}, \bibinfo{pages}{010402}
  (\bibinfo{year}{2005}).

\bibitem[{\citenamefont{Byers and Yang}(1961)}]{SCring}
\bibinfo{author}{\bibfnamefont{N.}~\bibnamefont{Byers}} \bibnamefont{and}
  \bibinfo{author}{\bibfnamefont{C.~N.} \bibnamefont{Yang}},
  \bibinfo{journal}{Phys. Rev. Lett.} \textbf{\bibinfo{volume}{7}},
  \bibinfo{pages}{46} (\bibinfo{year}{1961}).

\bibitem[{\citenamefont{B$\ddot{\rm{u}}$ttiker
  et~al.}(1983)\citenamefont{B$\ddot{\rm{u}}$ttiker, Imry, and
  Landauer}}]{ImryPersistent}
\bibinfo{author}{\bibfnamefont{M.}~\bibnamefont{B$\ddot{\rm{u}}$ttiker}},
  \bibinfo{author}{\bibfnamefont{Y.}~\bibnamefont{Imry}}, \bibnamefont{and}
  \bibinfo{author}{\bibfnamefont{R.}~\bibnamefont{Landauer}},
  \bibinfo{journal}{Phys. Lett.} \textbf{\bibinfo{volume}{96A}},
  \bibinfo{pages}{365} (\bibinfo{year}{1983}).

\bibitem[{\citenamefont{Loss and Maslov}(1995)}]{Loss_Maslov}
\bibinfo{author}{\bibfnamefont{D.}~\bibnamefont{Loss}} \bibnamefont{and}
  \bibinfo{author}{\bibfnamefont{D.~L.} \bibnamefont{Maslov}},
  \bibinfo{journal}{Phys. Rev. Lett.} \textbf{\bibinfo{volume}{74}},
  \bibinfo{pages}{178} (\bibinfo{year}{1995}).

\bibitem[{\citenamefont{Sch$\ddot{{\text{u}}}$tz
  et~al.}(2003)\citenamefont{Sch$\ddot{{\text{u}}}$tz, Kollar, and
  Kopietz}}]{Kopietz}
\bibinfo{author}{\bibfnamefont{F.}~\bibnamefont{Sch$\ddot{{\text{u}}}$tz}},
  \bibinfo{author}{\bibfnamefont{M.}~\bibnamefont{Kollar}}, \bibnamefont{and}
  \bibinfo{author}{\bibfnamefont{P.}~\bibnamefont{Kopietz}},
  \bibinfo{journal}{Phys. Rev. Lett.} \textbf{\bibinfo{volume}{91}},
  \bibinfo{pages}{017205} (\bibinfo{year}{2003}).

\bibitem[{\citenamefont{Chen and Sigrist}(2014)}]{sigrist}
\bibinfo{author}{\bibfnamefont{W.}~\bibnamefont{Chen}} \bibnamefont{and}
  \bibinfo{author}{\bibfnamefont{M.}~\bibnamefont{Sigrist}},
  \bibinfo{journal}{Phys. Rev. B} \textbf{\bibinfo{volume}{89}},
  \bibinfo{pages}{024511} (\bibinfo{year}{2014}).

\bibitem[{\citenamefont{Chen et~al.}(2013)\citenamefont{Chen, Horsch, and
  Manske}}]{wei2}
\bibinfo{author}{\bibfnamefont{W.}~\bibnamefont{Chen}},
  \bibinfo{author}{\bibfnamefont{P.}~\bibnamefont{Horsch}}, \bibnamefont{and}
  \bibinfo{author}{\bibfnamefont{D.}~\bibnamefont{Manske}},
  \bibinfo{journal}{Phys. Rev. B} \textbf{\bibinfo{volume}{87}},
  \bibinfo{pages}{214502} (\bibinfo{year}{2013}).

\bibitem[{\citenamefont{Schilling}()}]{Zurich}
\bibinfo{author}{\bibfnamefont{A.}~\bibnamefont{Schilling}},
  \bibinfo{note}{private communication}.

\bibitem[{\citenamefont{Kloss et~al.}(2010)\citenamefont{Kloss, Kreisel, and
  Kopietz}}]{kloss}
\bibinfo{author}{\bibfnamefont{T.}~\bibnamefont{Kloss}},
  \bibinfo{author}{\bibfnamefont{A.}~\bibnamefont{Kreisel}}, \bibnamefont{and}
  \bibinfo{author}{\bibfnamefont{P.}~\bibnamefont{Kopietz}},
  \bibinfo{journal}{Phys. Rev. B} \textbf{\bibinfo{volume}{81}},
  \bibinfo{pages}{104308} (\bibinfo{year}{2010}).

\bibitem[{\citenamefont{Snoke}(2006)}]{snoke}
\bibinfo{author}{\bibfnamefont{D.}~\bibnamefont{Snoke}},
  \bibinfo{journal}{Nature} \textbf{\bibinfo{volume}{443}},
  \bibinfo{pages}{403} (\bibinfo{year}{2006}).

\bibitem[{\citenamefont{Bunkov and Volovik}(2013, arXiv:1003.4889)}]{bunkov}
\bibinfo{author}{\bibfnamefont{Y.~M.} \bibnamefont{Bunkov}} \bibnamefont{and}
  \bibinfo{author}{\bibfnamefont{G.~E.} \bibnamefont{Volovik}},
  \emph{\bibinfo{title}{Novel Superfluids (Chapter IV); eds. K. H. Bennemann
  and J. B. Ketterson}} (\bibinfo{publisher}{Oxford University Press},
  \bibinfo{year}{2013, arXiv:1003.4889}).

\bibitem[{\citenamefont{Sonin}(2010)}]{sonin}
\bibinfo{author}{\bibfnamefont{E.~B.} \bibnamefont{Sonin}},
  \bibinfo{journal}{Adv. Phys.} \textbf{\bibinfo{volume}{59}},
  \bibinfo{pages}{181} (\bibinfo{year}{2010}).

\end{thebibliography}

\end{document}